\documentclass[11pt]{article}
\usepackage[dvips]{graphicx}
\usepackage{epsfig}
\usepackage{graphicx}
\usepackage{indentfirst}
\usepackage{amsmath}
\usepackage{amsfonts}
\usepackage{amssymb}

\begin{document}

\begin{center}
\huge{Bianchi type-I, type-III and Kantowski-Sachs solutions in $f(T)$ gravity} 
\end{center}

\begin{center}
{\small  M. E. Rodrigues $^{(a,b)}$}\footnote{E-mail
address: esialg@gmail.com}\,,
{\small  A. V. Kpadonou $^{(b,e)}$}\footnote{E-mail address: vkpadonou@gmail.com}\,,
{\small  F. Rahaman $^{(d)}$}\footnote{E-mail address: 
rahaman@iucaa.ernet.in}\,,\\
  {\small  P. J. Oliveira $^{(f)}$}\footnote{E-mail address: pauojoseo@gmail.com}\, and {\small    M. J. S. Houndjo $^{(b,c)}$}\footnote{E-mail address:
sthoundjo@yahoo.fr}

\vskip 4mm

(a) \ Faculdade de Ci\^{e}ncias Exatas e Tecnologia, Universidade Federal do Par\'{a} - Campus Universit\'{a}rio de Abaetetuba, CEP 68440-000, Abaetetuba, Par\'{a}, Brazil\\
(b) \ Institut de Math\'{e}matiques et de Sciences Physiques (IMSP) - 01 BP 613 Porto-Novo, B\'{e}nin\\
(c) \ Facult\'e des Sciences et Techniques de Natitingou - Universit\'e de Parakou - Natitingou - B\'{e}nin\\
(d) \ Department of Mathematics, Jadavpur University, Kolkata - 700032, India\\
(e)Ecole Normale Sup\'{e}rieure de Natitingou - Universit\'{e} de Parakou - B\'{e}nin\\
(f) Instituto Federal de Educa\c{c}\~{a}o, Ci\^{e}ncia e Tecnologia do Esp\'{i}rito Santo, 
 Campus Cachoeiro de Itapemirim - CEP 29.300-970 | Caixa Postal: 527 - Cachoeiro de Itapemirim - ES - Brazil
\vskip 2mm
\end{center}
\begin{abstract}
In the context of modified tele-parallel theory of gravity, we undertake cosmological anisotropic models and search for their solutions. Within a suitable choice of non-diagonal tetrads, the decoupled equations of motion are obtained for Bianchi-I, Bianchi-III and Kantowski-Sachs models, from which we obtain the correspondent solutions. By the way, energy density and pressures are also obtained, showing, as an important result, that our universe may live a quintessence like universe even still anisotropic models are considered.
\end{abstract}


\section{Introduction}
It is well known nowadays that modifying the law of gravity may lead to possible explanations for the acceleration mechanism of the our universe \cite{1-2dealvaro}. Various theories based on the modification of the law of gravity have been performed, among those, the modified version of the Tele-Parallel theory of gravity (TT), the so-called $f(T)$ gravity, where $T$ denotes the torsion scalar. The teleparallel gravity is a theory equivalent to the general relativity, but does not depend on the Levi-Civita's connection, as is the case of the GR. The TT and its modified version are based on the Weitzenbock connection where the curvature scalar vanishes whereas the torsion is different from zero.  Within this theory, various works have been developed and interesting results have been found, see \cite{gabriel,can}.\par
Note that the observational data are based on the assumption that the universe is homogeneous and isotropic for large scales. Indeed, the matter distribution in the universe is rigorously non-homogeneous and the propagation of the light may not be isotropic. In this way, it reasonable to look for at models able to approach this feature of the universe. This is the goal of this paper.\par
There are several metric models that traduce the anisotropic feature of the universe but we focus our attention on three of them, the Bianchi-I, Bianchi-III and Kantowski-Sachs. The motif of using these models is the fact that there their topologies, for large scales, may lead to the well known trivial FRW model within physical assumptions. An important point to be put out here is that the geometrical objects as the torsion and contorsion, which are indispensable for any calculus depend on the so-called tetrad. The tetrads play an important role and depending on their form, the theory may present some constraints on the gravitational action. In fact, for diagonal tetrads, constraints appears and the action is reduced to the TT one, with cosmological constant. However, when the non-diagonal tetrads are considered, the constraints disappear letting free the choice of the action.\par
Our goal is to search for the solutions of these models, say, finding the expression of the scale factor in the three space directions, within the modified TT. As consequence, through the dependence of the content of the universe of the geometrical part of the field equations, solutions are found for the energy density and the pressure, showing, as an interesting and important result, that still using anisotropic metric models, our universe may live a quintessence-like phase.\par
The paper is organised as follows: at the section $2$, we general equation of motion and write down the expression of the energy-momentum tensor within the assumption of inhomogeneous content of the universe. The section $3$ is devoted to the decoupling of the generalized equations following the pressures and the energy density. Still in this section solutions and comments are performed. The conclusion is presented at the section $4$.

\section{\large General equations of motion}\label{sec2}


We first define a metric for the geometry of the Weitzenbok spacetime as
\begin{equation}\label{el}
dS^{2}=g_{\mu\nu}dx^{\mu}dx^{\nu}\; ,
\end{equation} 
where $g_{\mu\nu}$ are the components of the metric which is symmetric and possess $10$ degrees of freedom. One can describe the theory in the spacetime or in the tangent space, which allows to rewrite the line element (\ref{el}) as follows 
\begin{eqnarray}
dS^{2} &=&g_{\mu\nu}dx^{\mu}dx^{\nu}=\eta_{ij}\theta^{i}\theta^{j}\label{1}\; ,\\
dx^{\mu}& =&e_{i}^{\;\;\mu}\theta^{i}\; , \; \theta^{i}=e^{i}_{\;\;\mu}dx^{\mu}\label{2}\; ,
\end{eqnarray} 
where $\eta_{ij}=diag[1,-1,-1,-1]$ and $e_{i}^{\;\;\mu}e^{i}_{\;\;\nu}=\delta^{\mu}_{\nu}$ or  $e_{i}^{\;\;\mu}e^{j}_{\;\;\mu}=\delta^{j}_{i}$. The square root of the metric determinant is given by  $\sqrt{-g}=\det{\left[e^{i}_{\;\;\mu}\right]}=e$ and the matrix $e^{a}_{\;\;\mu}$ are called tetrads and represent the dynamic fields of the theory. 
Note that the tetrads have a degree of freedom greater than the one of the metric, and this is showed in \cite{miao}, and we still have a possibility of non unequivocal choice \cite{daouda3}.
\par
We can define the Weitzenbock's connection as 
\begin{eqnarray}
\Gamma^{\alpha}_{\mu\nu}=e_{i}^{\;\;\alpha}\partial_{\nu}e^{i}_{\;\;\mu}=-e^{i}_{\;\;\mu}\partial_{\nu}e_{i}^{\;\;\alpha}\label{co}\; .
\end{eqnarray}
The main geometrical objects of the spacetime are constructed from this connection. The components of the tensor torsion are defined by the antisymmetric part of this connection
\begin{eqnarray}
T^{\alpha}_{\;\;\mu\nu}&=&\Gamma^{\alpha}_{\nu\mu}-\Gamma^{\alpha}_{\mu\nu}=e_{i}^{\;\;\alpha}\left(\partial_{\mu} e^{i}_{\;\;\nu}-\partial_{\nu} e^{i}_{\;\;\mu}\right)\label{tor}\;.
\end{eqnarray}
We also define the components of the so-called contorsion tensor as 
\begin{eqnarray}
K^{\mu\nu}_{\;\;\;\;\alpha}&=&-\frac{1}{2}\left(T^{\mu\nu}_{\;\;\;\;\alpha}-T^{\nu\mu}_{\;\;\;\;\alpha}-T_{\alpha}^{\;\;\mu\nu}\right)\label{contor}\; .
\end{eqnarray}
In order to make more clear the definition of the scalar equivalent to the curvature scalar of RG, we first define a new tensor $S_{\alpha}^{\;\;\mu\nu}$, constructed from the components of the torsion and contorsion tensors as
\begin{eqnarray}
S_{\alpha}^{\;\;\mu\nu}&=&\frac{1}{2}\left( K_{\;\;\;\;\alpha}^{\mu\nu}+\delta^{\mu}_{\alpha}T^{\beta\nu}_{\;\;\;\;\beta}-\delta^{\nu}_{\alpha}T^{\beta\mu}_{\;\;\;\;\beta}\right)\label{s}\;.
\end{eqnarray}
Now, we are able to construct a contraction which is equivalent to the scalar curvature in GR. We define then the torsion scalar as
\begin{eqnarray}
T=T^{\alpha}_{\;\;\mu\nu}S^{\;\;\mu\nu}_{\alpha}\label{te}\; .
\end{eqnarray}
As $f(R)$ gravity generalizes the GR, we also define the action of  $f(T)$ gravity  as the generalization of the teleparallel, being the continue sum     
\begin{eqnarray}\label{action}
S=\int d^4x\,e\left[f(T)+\mathcal{L}_{Matter}\right]\;,
\end{eqnarray}
where $f(T)$ is an algebraic function of the torsion scalar $T$. Making the functional variation of the action  (\ref{action}) with respect to the tetrads, we get the following equations of motion \cite{barrow,daouda2}
\begin{eqnarray}
S^{\;\;\nu\rho}_{\mu}\partial_{\rho}Tf_{TT}+\left[e^{-1}e^{i}_{\mu}\partial_{\rho}\left(ee^{\;\;\alpha}_{i}S^{\;\;\nu\rho}_{\alpha}\right)+T^{\alpha}_{\;\;\lambda\mu}S^{\;\;\nu\lambda}_{\alpha}\right]f_{T}+\frac{1}{4}\delta^{\nu}_{\mu}f=4\pi\mathcal{T}^{\nu}_{\mu}\label{em}\; ,
\end{eqnarray}
where $\mathcal{T}^{\nu}_{\mu}$ is the energy momentum tensor, $f_{T}=d f(T)/d T$ and $f_{TT}=d^{2} f(T)/dT^{2}$. By setting $f(T)=T-2\Lambda$,  the equations of motion (\ref{em}) are the same as that of the Teleparallel theory with a cosmological constant, and this is dynamically equivalent to the GR.  These equations clearly depend on the choice made for the set of tetrads \cite{cemsinan}. 
\par
The contribution of the interaction with the matter fields is given by the energy momentum tensor which, is this case, is defined as 
\begin{eqnarray}
\mathcal{T}^{\,\nu}_{\mu}=\left(\rho+p_t\right)u_{\mu}u^{\nu}-p_t \delta^{\nu}_{\mu}+\left(p_r-p_t\right)v_{\mu}v^{\nu}\label{tem}\; ,
\end{eqnarray}
where $u^{\mu}$ is the four-velocity, $v^{\mu}$ a unitary space-like vector in the radial direction, $\rho$ the energy density, $p_r$ the pressure in the direction of $v^{\mu}$ (radial pressure) and $p_t$  the pressure orthogonal to $v_\mu$ (tangential pressure). This tensor characterizes an anisotropic fluid. 


\section{Anisotropic models}\label{sec3}

In this section, we will study the equations of motion coming from the metric of Bianchi type-I, type-III and Kantowski-Sachs, first in a unified way and with a choice of two sets of tetrads, the diagonal and the non-diagonal.

\subsection{Diagonal tetrads}\label{subsec3.1}
In this subsection we will study the case of a choice of a set of diagonal tetrads for some inhomogeneous models.  We can describe the models of Bianchi type-I, type-III and  Kantowski-Sachs (KS) through the metric 
\begin{equation}
dS^2=dt^2-A^2(t)dr^2-B^2(t)\left[d\theta^2+K_j^2(\theta)d\phi^2\right]\;,
\end{equation}
where $j=1,2,3$, $K_1(\theta)=\theta,K_2(\theta)=\sin\theta$ and  $K_3(\theta)=\sinh\theta$ are the models of  Bianchi type-I, Kantowski-Sachs and  Bianchi type-III, respectively. Let us start choosing a set of diagonal tetrads 
\begin{eqnarray}\label{td}
\left[e^{a}_{\;\;\mu}\right]=diag\left[1,A,B,BK_j(\theta)\right]\;.
\end{eqnarray}
The determinant of this matrix is  $e=AB^2K_j(\theta)$. The components of the torsion tensor  (\ref{tor}), for the choice of tetrads  (\ref{td}), are given by 
\begin{eqnarray}\label{tord}
T^{1}_{\;\;01}=\frac{\dot{A}}{A}\,,\,T^{2}_{\;\;02}=T^{3}_{\;\;03}=\frac{\dot{B}}{B}\,,\,T^{3}_{\;\;23}=\frac{1}{K_j}\frac{dK_j}{d\theta}\;,
\end{eqnarray}      
those the correspond the contorsion  (\ref{contor}) components are 
\begin{eqnarray}\label{contord}
K^{01}_{\;\;\;\;1}=\frac{\dot{A}}{A}\,,\,K^{02}_{\;\;\;\;2}=K^{03}_{\;\;\;\;3}=\frac{\dot{B}}{B}\,,\,K^{32}_{\;\;\;\;3}=\frac{1}{B^2K_j}\frac{dK_j}{d\theta}\;,
\end{eqnarray}      
and those of the tensor $S_{\alpha}^{\;\;\mu\nu}$, in (\ref{s}),  are given by \begin{eqnarray}\label{sd}
S_{1}^{\;\;10}=\frac{\dot{B}}{B}\,,\,S_{0}^{\;\;20}=S_{1}^{\;\;21}=\frac{1}{2B^2K_j}\frac{dK_j}{d\theta}\,,\,S_{2}^{\;\;20}=S_{3}^{\;\;30}=\frac{1}{2}\left(\frac{\dot{A}}{A}+\frac{\dot{B}}{B}\right)\;.
\end{eqnarray}
Making use of the components (\ref{tord}) and (\ref{sd}), the torsion scalar (\ref{te}) is given by 
\begin{eqnarray}\label{ted}
T=-2\left(\frac{\dot{B}}{B}\right)^2-4\frac{\dot{A}\dot{B}}{AB}\;.
\end{eqnarray}

The equations  of motion  (\ref{em}) are
\begin{eqnarray}
4\pi\rho&=&\frac{f}{4}+f_{T}\left[\left(\frac{\dot{B}}{B}\right)^2+2\frac{\dot{A}\dot{B}}{AB}+\frac{k}{2B^2}\right]\;,\label{densd}\\
-4\pi p_r&=&\frac{\dot{B}}{B}\dot{T}f_{TT}+\frac{f}{4}+f_{T}\left[\frac{\ddot{B}}{B}+\left(\frac{\dot{B}}{B}\right)^2+\frac{\dot{A}\dot{B}}{AB}+\frac{k}{2B^2}\right]\;,\label{prd}\\
-4\pi p_t&=&\frac{1}{2}\left(\frac{\dot{A}}{A}+\frac{\dot{B}}{B}\right)\dot{T}f_{TT}+\frac{f}{4}+\frac{f_{T}}{2}\left[\frac{\ddot{A}}{A}+\frac{\ddot{B}}{B}+\left(\frac{\dot{B}}{B}\right)^2+3\frac{\dot{A}\dot{B}}{AB}\right]\;,\label{ptd}\\
\frac{\dot{T}(dK_j/d\theta)}{2B^2K_j}f_{TT}&=&0\label{constraint1}\;,
\end{eqnarray}
where $k=-K_j^{-1}(d^2K_j/d\theta^2)$. For the Kantowski-Sachs model $k=+1$, for Bianchi type-I $k=0$, and for Bianchi type-III $k=-1$. This classification stems from the scalar curvature of the spatial hyper-surface $^{(3)}R=2k/B^2(t)$, in the case of the GR  \cite{paulo1}. 
\par
The constraint equation (\ref{constraint1}) imposes to the function $f(T)$ to be linear, as in the Teleparallel Theory (TT), and we can choose it as  $f(T)=T-2\Lambda$, where  $\Lambda$ is the cosmological constant. Hence, the equations  (\ref{densd})-(\ref{ptd}) become
\begin{eqnarray}
\left(\frac{\dot{B}}{B}\right)^2+2\frac{\dot{A}\dot{B}}{AB}+\frac{k}{B^2}&=&\Lambda+8\pi\rho\label{densd1}\,,\\
2\frac{\ddot{B}}{B}+\left(\frac{\dot{B}}{B}\right)^2+\frac{k}{B^2}&=&\Lambda-8\pi p_r\label{prd1}\,,\\
\frac{\ddot{A}}{A}+\frac{\ddot{B}}{B}+\frac{\dot{A}\dot{B}}{AB}&=&\Lambda-8\pi p_t\label{ptd1}\,.
\end{eqnarray} 
These equations are the same as those of the GR \cite{paulo2}, and then, all the known solutions for these models can be regained. Once again, this is not surprising, since the TT is dynamically equivalent to the GR and the equations of motion must be identical. We can also observe that the particular case $A(t)=B(t)=a(t)$ and $p_r=p_t$, for the model of Bianchi type-I ($k=0$), we regain the flat FLRW universe in  (\ref{densd1})-(\ref{ptd1}), with $a(t)$ being the scale factor.  

In the next section, we will study a choice of a set of non-diagonal tetrads.

\subsection{Non-diagonal tetrads}\label{subsec3.2}


In this subsection we will study the inhomogeneous models, with a set of non-diagonal tetrads, similar to the static case \cite{daouda3}, or in the cosmological case \cite{daouda2}. Our goal here is to get some slight modification in the equations of motion, coming from a set of non-diagonal tetrads, and compare the results with the diagonal case. This can lead in general to some understanding for the use of anisotropic models.
\par
By choosing the set of non-diagonal tetrads, 
\begin{eqnarray}\label{tnd}
\left[e^{a}_{\;\;\mu}\right]=\left[\begin{array}{cccc}
1 & 0 & 0 & 0\\
0 & A\sin\theta\cos\phi & B\cos\theta\cos\phi & -BK_j(\theta)\sin\phi\\
0 & A\sin\theta\sin\phi & B\cos\theta\sin\phi & BK_j(\theta)\cos\phi\\
0 & A\cos\theta & -B\sin\theta & 0
\end{array}\right]\;,
\end{eqnarray} 
we get the determinant $e=AB^2K_j$. The components of the torsion  (\ref{tor}), contorsion (\ref{contor}) and tensor $S_{\alpha}^{\;\;\mu\nu}$ (\ref{s}), are given by 
\begin{eqnarray}
T^{1}_{\;\;01}&=&\frac{\dot{A}}{A}\,,\,T^{2}_{\;\;02}=T^{3}_{\;\;03}=\frac{\dot{B}}{B} \,,\,T^{2}_{\;\;21}=\frac{A}{B} \,,\,T^{3}_{\;\;31}=\frac{A\sin\theta}{BK_j}\,,\,T^{3}_{\;\;23}=\frac{(dK_j/d\theta)-\cos\theta}{K_j}\label{tornd}\;,\\
K^{01}_{\;\;\;\;1}&=&\frac{\dot{A}}{A}\,,\,K^{02}_{\;\;\;\;2}=K^{03}_{\;\;\;\;3}=\frac{\dot{B}}{B} \,,\,K^{12}_{\;\;\;\;2}=\frac{1}{AB} \,,\,K^{13}_{\;\;\;\;3}=\frac{\sin\theta}{ABK_j}\,,\,K^{32}_{\;\;\;\;3}=\frac{(dK_j/d\theta)-\cos\theta}{B^2K_j}\;,\label{contornd}\\
&&\left\{\begin{array}{ll}\label{snd}
S_{0}^{\;\;01}=\frac{K_j+\sin\theta}{2ABK_j} \,,\,S_{0}^{\;\;20}=S_{1}^{\;\;21}=\frac{(dK_j/d\theta)-\cos\theta}{2B^2K_j} \,,\,S_{1}^{\;\;10}=\frac{\dot{B}}{B} \,,\\
S_{2}^{\;\;20}=S_{3}^{\;\;30}=\frac{1}{2}\left(\frac{\dot{A}}{A}+\frac{\dot{B}}{B}\right) \,,\,S_{2}^{\;\;21}= \frac{\sin\theta}{2ABK_j}\,,\,S_{3}^{\;\;31}=\frac{1}{2AB}\,.
\end{array}\right.
\end{eqnarray}
The torsion scalar  (\ref{te}), from the components (\ref{tornd}) and (\ref{snd}), is given by 
\begin{equation}\label{tend}
T=-2\left(\frac{\dot{B}}{B}\right)^2-4\frac{\dot{A}\dot{B}}{AB}+\frac{2\sin\theta}{B^2K_j}
\end{equation}
The equations of motion  (\ref{em}) become 
\begin{eqnarray}
4\pi\rho&=&-\left[\frac{(dK_j/d\theta)-\cos\theta}{2B^2K_j}\right]\frac{dT}{d\theta}f_{TT}+\frac{f}{4}+f_{T}\left[\left(\frac{\dot{B}}{B}\right)^2+2\frac{\dot{A}\dot{B}}{AB}+\frac{k-(\sin\theta/K_j)}{2B^2}\right]\;,\label{densnd}\\
-4\pi p_r&=&\frac{\dot{B}}{B}\dot{T}f_{TT}+\frac{f}{4}+f_{T}\left[\frac{\ddot{B}}{B}+\left(\frac{\dot{B}}{B}\right)^2+\frac{\dot{A}\dot{B}}{AB}+\frac{k-(\sin\theta/K_j)}{2B^2}\right]\;,\label{prnd}\\
-4\pi p_t&=&\frac{1}{2}\left(\frac{\dot{A}}{A}+\frac{\dot{B}}{B}\right)\dot{T}f_{TT}+\frac{f}{4}+\frac{f_{T}}{2}\left[\frac{\ddot{A}}{A}+\frac{\ddot{B}}{B}+\left(\frac{\dot{B}}{B}\right)^2+3\frac{\dot{A}\dot{B}}{AB}-\frac{\sin\theta}{B^2K_j}\right]\;,\label{ptnd}\\
&&\frac{K_j+\sin\theta}{2ABK_j}\dot{T}f_{TT}=0\,,\label{constraint2}
\end{eqnarray}
\begin{eqnarray}
&&\frac{(dK_j/d\theta)-\cos\theta}{2B^2K_j}\dot{T}f_{TT}=0\label{constraint3}\;,\\
&&\frac{\sin\theta}{2ABK_j}\frac{dT}{d\theta}f_{TT}=0\,,\label{constraint4}\\
&&\frac{1}{2}\left(\frac{\dot{A}}{A}+\frac{\dot{B}}{B}\right)\frac{dT}{d\theta}f_{TT}=0\label{constraint5}\;.
\end{eqnarray}
Also here, the constraint equations  (\ref{constraint2})-(\ref{constraint5}) impose to the function $f(T)$ to be linear or constant torsion.
\par
By choosing $f(T)=T-2\Lambda$, the equations of motion  (\ref{densnd})-(\ref{ptnd}) turn into 
\begin{eqnarray}
\left(\frac{\dot{B}}{B}\right)^2+2\frac{\dot{A}\dot{B}}{AB}+\frac{k}{B^2}&=&\Lambda+8\pi\rho\label{densnd1}\,,\\
2\frac{\ddot{B}}{B}+\left(\frac{\dot{B}}{B}\right)^2+\frac{k}{B^2}&=&\Lambda-8\pi p_r\label{prnd1}\,,\\
\frac{\ddot{A}}{A}+\frac{\ddot{B}}{B}+\frac{\dot{A}\dot{B}}{AB}&=&\Lambda-8\pi p_t\label{ptnd1}\,.
\end{eqnarray}

We can define  the average volume as $V(t)=A(t)B^2(t)$. Supposing first an exponential expansion $V(t)=v_0e^{3H_0 t}$, with  $v_0,H_0\in\Re^{+}$, one gets 
\begin{equation}
A(t)=\frac{v_0e^{3H_0 t}}{B^2(t)}\label{a1}\,.
\end{equation}
Now, defining the average Hubble's parameter as
\begin{equation}
 H(t)=\frac{1}{3}\frac{\dot{V}}{V}=\frac{1}{3}\left(\frac{\dot{A}}{A}+2\frac{\dot{B}}{B}\right)\label{h}\,,
\end{equation}
one has  $H(t)=H_0$. By defining a deceleration factor $q=[(d/dt)(1/H(t))-1]$, one gets  $q=-1$ which yields an accelerated universe. We also define  $H_1=\dot{A}/A,H_{2,3}=\dot{B}/B$, and the anisotropic parameter of the expansion as
\begin{equation}\label{del}
\Delta (t)= \frac{1}{3}\sum^{3}_{i=1}\left(\frac{H_i}{H}-1\right)^2=\frac{2}{9H^2}\left(\frac{\dot{A}}{A}-\frac{\dot{B}}{B}\right)^2\,.
\end{equation}
Making similar considerations to those of Adhav et al \cite{adhav}, we can simplify the equations, using $p_r=\omega(t)\rho(t)$ and  $p_t=[\omega(t)+\delta(t)]\rho(t)$, with 
\begin{equation}
\rho(t)=\frac{k}{8\pi\delta(t)B^2(t)}\,.\label{dens1}
\end{equation} 

Then, we can solve the equations of motion (\ref{densnd1})-(\ref{ptnd1}), setting  $B(t)=\exp\left[b(t)\right]$, which yields 
\begin{eqnarray}
6H_0\dot{b}+ke^{-2b}-3\dot{b}^2-\frac{ke^{-2b}}{\delta}-\Lambda&=&0\label{eq1}\,,\\
3\dot{b}^2+2\ddot{b}+ke^{-2b}+\frac{ke^{-2b}}{\delta}\omega-\Lambda&=&0\label{eq2}\,,\\
9H_0^2+3\dot{b}^2-9H_0\dot{b}-\ddot{b}+ke^{-2b}+\frac{ke^{-2b}}{\delta}\omega-\Lambda&=&0\label{eq3}\,.
\end{eqnarray}
Subtracting (\ref{eq2}) from  (\ref{eq3}) one gets 
\begin{equation}
3\ddot{b}+9H_0\dot{b}-9H_0^2=0\,\label{b-1}
\end{equation}
leading to  $b(t)=H_0 t-(b_0e^{-3H_0 t}/3H_0)+b_1$, with  $b_0\in\Re$. The solution is then given by 
\begin{eqnarray}
A(t)=\frac{v_0}{b_2^2}\exp\left(H_0 t+\frac{2b_0}{3H_0}e^{-3H_0t}\right)\,,\,B(t)=b_2\exp\left(H_0t-\frac{b_0}{3H_0}e^{-3H_0t}\right)\,.\label{sol1}
\end{eqnarray}
The anisotropic parameter of the expansion (\ref{del}) is given by 
\begin{equation}
\Delta(t)=\frac{2c_2^2}{k^2}e^{-6kt}\label{del1}\,.
\end{equation}
We can now isolate  $\delta(t)$ in the equation  (\ref{eq1}), getting  
\begin{eqnarray}
\delta(t)=\frac{\exp\left(4kt+\frac{2s_2}{3k}e^{-3kt}\right)}{c_3^2(3k^2e^{6kt}-3c_2^2-\Lambda e^{6kt})}\label{delta1}\,.
\end{eqnarray} 
We also isolate  $\omega(t)$ in  (\ref{eq2}), obtaining 
\begin{eqnarray}
\omega(t)=-\frac{3k^2e^{6kt}+3c_2^2-\Lambda e^{6kt}}{3k^2e^{6kt}-3c_2^2-\Lambda e^{6kt}}\label{omega1}\,.
\end{eqnarray} 
We present the curve that traduces the evolution of this parameter versus $t$ for some values of the input parameters. Hence, we see that, setting $k^2=1$, $\Lambda=0.01$ and $c_2^2=0.1$, the parameter $\omega$ tends to $-1$; see Fig. $1$. \par
The density (\ref{dens1}) becomes 
\begin{eqnarray}
\rho(t)=\frac{3k^2-3c_2^2e^{-6kt}-\Lambda}{8\pi}\label{dens2}\,.
\end{eqnarray} 

Now we can test the physical criteria for this solution, as we have in \cite{collins}. The following criteria have to be obeyed: a) density  $\rho(t)$ always positive ; b) volume $V(t)$ going to infinity when  $t\rightarrow\infty$; c) the anisotropic parameter of the expansion $\Delta(t)$ must tend to zero when  $t\rightarrow\infty$. For  $k^2>c_2^2+\Lambda/3$ the density  (\ref{dens2}) is always positive, then, satisfies the criterion  a). In the limit  $t\rightarrow\infty$, one gets      
\begin{eqnarray}
\Delta(t)\rightarrow0\,,\,V(t)\rightarrow\infty\,,\,\rho(r)\rightarrow\frac{3k^2-\Lambda}{8\pi}\,,\,\delta(t)\rightarrow 0\,,\,\omega(t)\rightarrow -1\,.
\end{eqnarray}
Note that the solution is physical, the anisotropy disappears, the density is always positive, the volume tends to infinity and the model tends to the $\Lambda$CDM one, with $\omega=-1$.\par
We also consider a power type expansion, i.e,  $V(t)=c_1 t^{3n}$, with  $c_1,n\in\Re^{+}$. Then, we have $A(t)=c_1 t^{3n}/B^2(t)$ and  $H(t)=n/t$. In this case, considering  (\ref{dens1}) and $B=\exp(b)$, the equations  (\ref{densnd1})-(\ref{ptnd1}), for  $\Lambda=0$, yield 
\begin{eqnarray}
3\dot{b}^2-\frac{6n}{t}\dot{b}+\frac{e^{-2b}}{\delta}&=&0\label{eq1-1}\,\\
2\ddot{b}+3\dot{b}^2+\frac{e^{-2b}}{\delta}\omega&=&0\label{eq2-1}\,\\
-\ddot{b}+3\dot{b}^2-\frac{9n}{t}\dot{b}-\frac{3n(1-3n)}{t^2}+\frac{e^{-2b}}{\delta}\omega&=&0\label{eq3-1}\,.
\end{eqnarray}
Subtracting  (\ref{eq2-1}) from  (\ref{eq3-1}), one gets 
\begin{equation}
t^2\ddot{b}+3nt\dot{b}+n(1-3n)=0\,,
\end{equation}
which, after integration, yields  $b(t)=\log(t^n)+[c_2t^{1-3n}/(1-3n)]+b_0$. The solution is given by  
\begin{eqnarray}
A(t)=\frac{c_1}{c_2^2}t^n \exp\left(\frac{-2t^{1-3n}}{1-3n}\right)\,,\,B(t)=c_2t^n\exp\left(\frac{2t^{1-3n}}{1-3n}\right)\label{sol2}
\end{eqnarray}
The anisotropic parameter of the expansion (\ref{del}) is given by 
\begin{equation}
\Delta(t)=\frac{2t^{2-3n}}{n^2}\label{del2}\,.
\end{equation}
We can now isolate  $\delta(t)$ in the equation (\ref{eq1-1}), getting 
\begin{eqnarray}
\delta(t)=\frac{t^{2+4n}\exp\left(\frac{-2t^{1-3n}}{1-3n}\right)}{3c_2^2(n^2t^{6n}-t^2)}\label{delta2}\,.
\end{eqnarray} 
Also, isolating   $\omega(t)$ in the equation  (\ref{eq2-1}), one gets 
\begin{eqnarray}
\omega(t)=\frac{3t^2+n(3n-2)t^{6n}}{3(t^2-n^2t^{6n})}\label{omega2}\,.
\end{eqnarray} 
In this case, we plot the evolution of $\omega$ versus time for two different cases. The first one, Fig. $2$, shows that the parameter $\omega$ tends to $-1/3$ only when $n$ is very closed to $1$. On other hand, we see that, for $n>1$, more precisely $n=2$ (as an example), $\omega$ does not tend to $-1/3$, but rather $0.66$; see Fig. $3$. Therefore, we conclude that the parameter $n$ is responsible for quintessence era.\par

The density  (\ref{dens1}) becomes 
\begin{eqnarray}
\rho(t)=\frac{3(\frac{n^2}{t^2}-t^{-6n})}{8\pi}\label{dens3}\,.
\end{eqnarray} 

Let us now test the physical criteria for this solution. For  $t>n\geq1$ the density  (\ref{dens3}) is always positive, and then, satisfies the criterion  a). For  $t\rightarrow\infty$, there is a dependence on the value of  $n$. There is a  quintessence-like behaviour, with  $\omega\in(-1,0)$. In the particular case,  $n=1$, we have       
\begin{eqnarray}
\Delta(t)\rightarrow0\,,\,V(t)\rightarrow\infty\,,\,\rho(r)\rightarrow 0\,,\,\delta(t)\rightarrow 0\,,\,\omega(t)\rightarrow -\frac{1}{3}\,.
\end{eqnarray}
Note that the solution is physical, the anisotropy disappears, the energy density is always positive, the volume tends to infinity and the model tends to a quintessence-like one.

\begin{figure}
\begin{center}
\includegraphics[angle=0.5, width=1\textwidth]{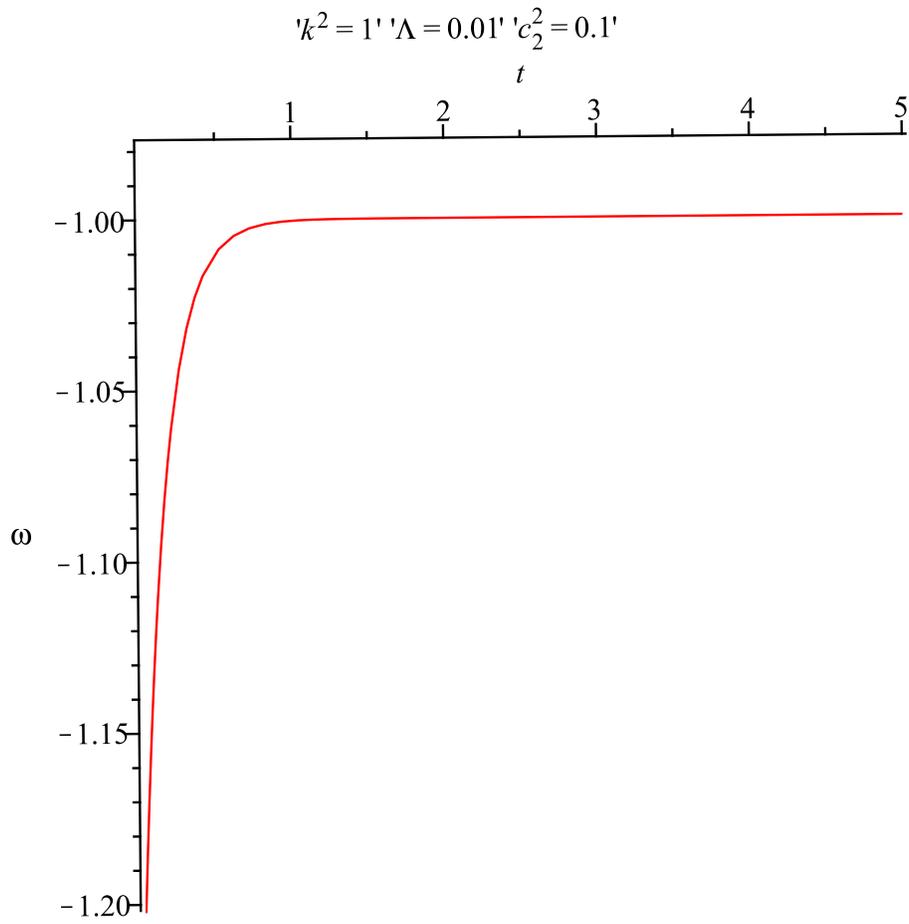}
\end{center}
\caption{\label{fig1}  This figure point out the evolution of the parameter $\omega$ in terms of $t$ for  $k^2=1$, $\Lambda=0.01$ and $c_2^2=0.1$.}
\end{figure}
\begin{figure}

\begin{center}
\includegraphics[angle=0.5, width=1\textwidth]{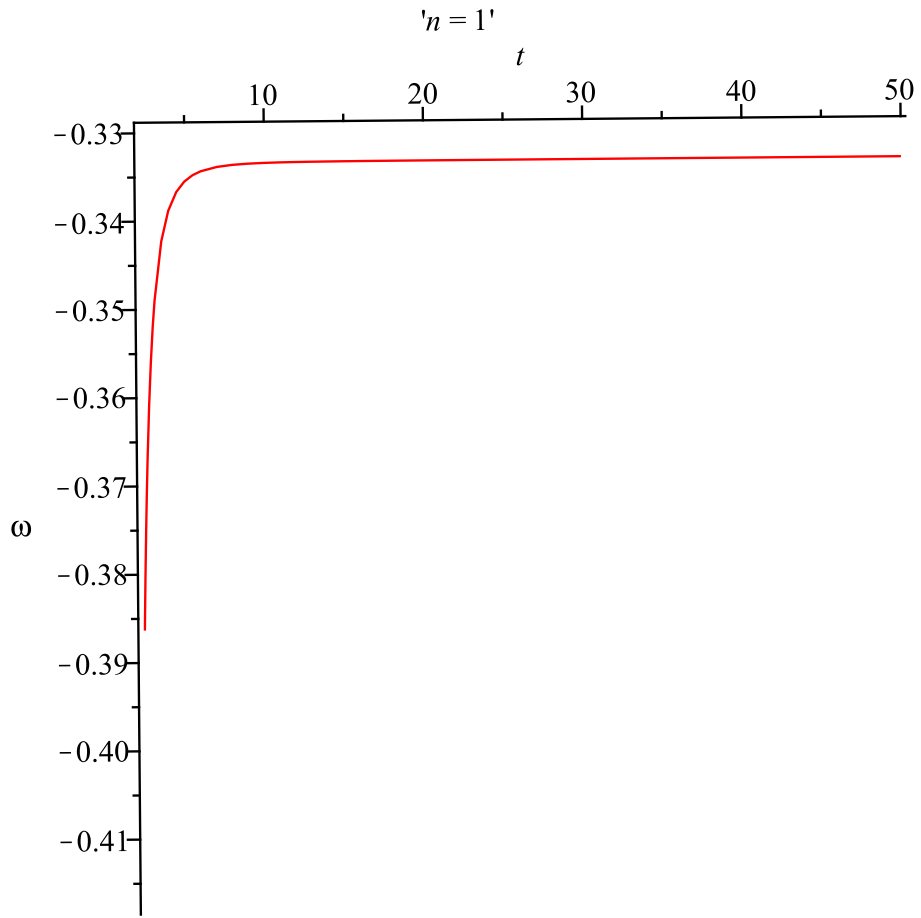}
\end{center}
\caption{\label{fig2} The figure shows the evolution of the parameter $\omega$ as the time evolves, for $n=1$. }
\end{figure}

\begin{figure}
\begin{center}
\includegraphics[angle=0.5, width=1\textwidth]{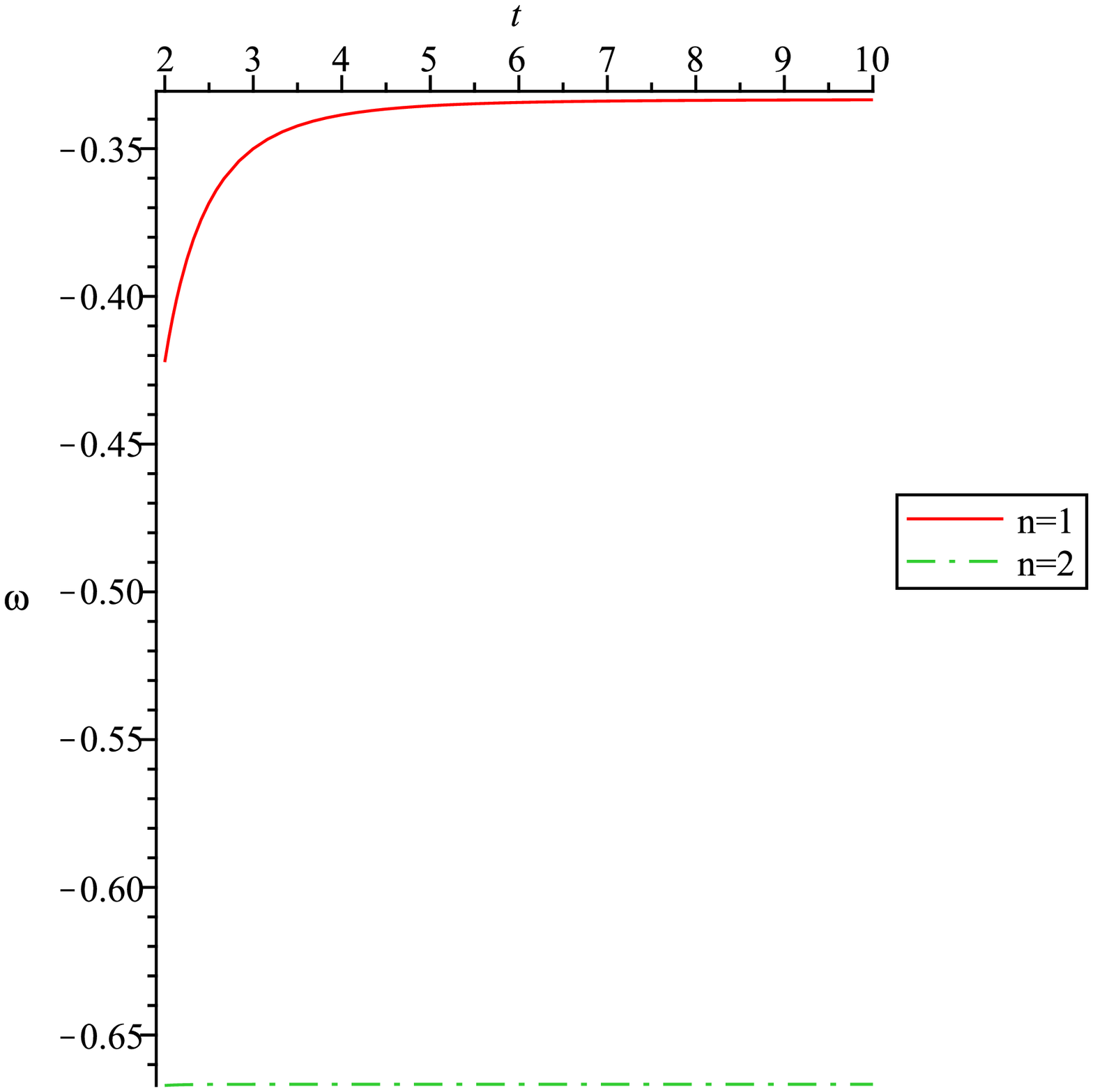}
\end{center}
\caption{\label{fig3}  This figure presents two curves, one for $n=1$ and the second for $n=2$, both pointing out evolutions of the parameter $\omega$. }
\end{figure}

\section{Conclusion}\label{sec4} 
In this paper, we undertake the modified tele-parallel theory, $f(T)$ gravity, where $T$ denotes the torsion scalar. In order to approach the real feature of the universe, due to the fact that the universe is not rigorously isotropic and homogeneous, we look for three interesting and realistic metric models, the Bianchi-I, Bianchi-III and Kantowski-Sachs models, and tried to solve the equations of motion. We focus our attention on non-diagonal tetrads where the gravitational  action is free from the constraint of recovering the TT. In this way, we found solutions for the three space directions, and through the equations of motion, the expressions of the energy density and pressures are obtained. An analysis of these expressions showed, as an interesting result that, even considering  that our universe is isotropic, it may live in a quintessence-like phase.

\newpage


\vspace{0,25cm}


\end{document}